# Seismic vulnerability analysis of moderate seismicity areas using in situ experimental techniques: from the building to the city scale –Application to Grenoble and Pointe-à-Pitre (France)

MICHEL Clotaire
*Laboratoire d'Informatique et Mécanique Appliquées à la Construction (IMAC),
Ecole Polytechnique Fédérale de Lausanne (EPFL)
Laboratoire de Géophysique Interne et de Tectonophysique (LGIT), CNRS,
Université Joseph Fourier Grenoble*

GUEGUEN, Philippe
*Laboratoire de Géophysique Interne et de Tectonophysique (LGIT), LCPC,
CNRS, Université Joseph Fourier Grenoble*

## Abstract

Seismic vulnerability analysis of existing buildings requires basic information on their structural behaviour. The ambient vibrations of buildings and the modal parameters (frequencies, damping ration and modal shapes) that can be extracted from them naturally include the geometry and quality of material in the linear elastic part of their behaviour. The aim of this work is to use this modal information to help the vulnerability assessment.

A linear dynamic modal model based on experimental modal parameters is proposed and the fragility curve corresponding to the damage state "Slight" is built using this model and a simple formula is proposed. This curve is particularly interesting in moderate seismic areas.

This methodology is applied to the Grenoble City where ambient vibrations have been recorded in 61 buildings of various types and to the Pointe-à-Pitre City with 7 study-buildings. The fragility curves are developed using the aforementioned methodology. The seismic risk of the study-buildings is discussed by performing seismic scenarios.

## Introduction

Seismic vulnerability analysis of existing buildings requires basic information on their structural behaviour. In the current methods such as HAZUS [1] or Risk-UE [2], these parameters are collected by visual screening or estimated among generic values of physical parameters.

Even for more complete diagnosis, the experts have to deal with the lack of structural plans, the unknown quality of material, ageing and damaging of the structure. The ambient vibrations of buildings and the modal parameters (frequencies, damping ratios and modal shapes) that can be extracted from them naturally include all these parameters in the linear elastic part of their behaviour. The aim of this work is to use this modal information to help the vulnerability assessment.

Moreover, there is a need in earthquake engineering methods to better predict the first damage grades with rather simple methods. In the case of countries with a moderate seismicity, there is a particular interest to quickly assess or to predict if an earthquake can cause slight damage or not. In that way, elastic models are still valuable information.

In this paper we present a method of assessment of the fragility of buildings concerning the first damage grade by using experimental modal parameters obtained under ambient vibrations. The suggested method is applied to the city of Grenoble (France) and buildings of Pointe-à-Pitre (Guadeloupe, France) to be validated.

## Ambient vibrations and modal analysis

A permanent oscillatory motion is affecting civil engineering structures caused by natural (ocean, atmosphere…) and anthropogenic (traffic, industries…) vibrations of the ground, wind and internal load. It has been shown [3] that the modes of vibrations of the buildings where the same on several order of magnitude of vibration amplitude. Moreover, Dunand et al. [4] showed on Californian buildings permanently instrumented that the resonance frequencies of buildings were decreasing at worst of 20% without serious damage. Consequently the modal information contained in ambient vibrations recorded in situ can be used (carefully) for earthquake engineering purposes.

Operational Modal Analysis provides a large amount of techniques to extract modal parameters (resonance frequencies, modal shapes and damping ratios) from ambient vibration recordings, especially in civil engineering structures, assuming a white noise input. In this paper the selected technique is the Frequency Domain

Decomposition [5] for its simplicity and the quality of its results (Fig. 1). Moreover it performs the best of simultaneous recordings. The response of civil engineering structures is not so simple than for example well controlled mechanics pieces. Therefore, modal analysis of such signals should not be made as a black box, at least in a first phase. Unless it just performs signal processing, the FDD technique allows a real decomposition of the modes and can give the damping ratios in case of qualitative data. The idea of this method is to compute the power spectral density matrices of simultaneous recordings and to diagonalize these matrices. The modes are then classified on the eigenvalues with respect to their energy in the signal (Fig. 1a). Peaks in the eigenvalues denote the presence of modes with modal shape on the corresponding eigenvector (Fig. 1b). The modes of the structures are generally carried by the first eigenvalue but even a mode with little energy can be identified. The (automatic) selection of the bell corresponding to a mode allows to identify more precisely the frequency and the damping ratio [6].

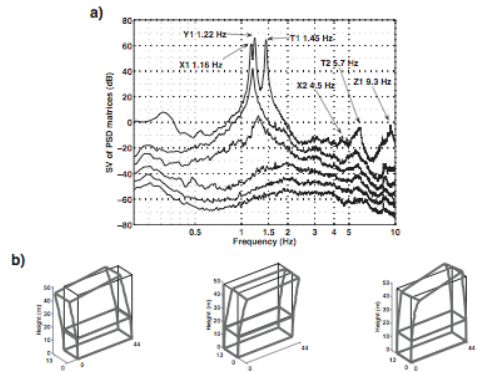

*Figure 1 – Modal Analysis using FDD of the Grenoble City Hall building. Eigenvalues of the PSD matrices (a). Modal shapes for the three first peaks (b).*

## Modelling

The modal parameters extracted from ambient vibrations are only valid in the linear elastic part of behaviour of the building. Assuming a 1D linear lumped-mass modelling, an analytic solution of the response $\{U(t)\}$ of each floor is available (Duhamel integral) [7]:

$$\{U(t)\} = [\Phi]\{y(t)\} + U_s(t) \quad (1)$$

with $\forall j \in [1,N]$ 
$$y_j(t) = \frac{-p_j}{\omega'} \int_0^t U_s''(\tau) e^{-\xi_j \omega_j (t-\tau)} \sin(\omega'(t-\tau)) d\tau \quad (2)$$

$$\omega_j'^2 = \omega_j^2(1-\xi_j^2) \quad \text{and} \quad p_j = \frac{\{\Phi_j\}^T[M]\{1\}}{\{\Phi_j\}^T[M]\{\Phi_j\}} = \frac{\sum_{i=1}^{N}\Phi_{ij}}{\sum_{i=1}^{N}\Phi_{ij}^2} \tag{3}$$

This formula only depends on the input motion Us(t), the modal parameters ([Φ] the modal shapes, {ω} the resonance angular frequencies and {ξ} the damping ratios). Assuming that the mass is constant at each floor, the participation factor of jth mode $p_j$ itself only depends on the modal shapes (Eq. 3). In that way, the modelling can only be based on experimental modal parameters and does not need any other assumption. This model has been validated on buildings in Grenoble [8].

## Fragility estimation

In order to predict damage, the interstory drift ratio has been chosen as damage index as in HAZUS [1]. This parameter is a natural output of the modelling presented above and standard values of these parameters are given in HAZUS [1] for different types of buildings and damage grades. As a linear model is used, only the first damage grade is considered. The interstory drifts at Slight damage used in this study are summarized on table 1.

*Table 1 – Interstory drift ratios at Slight damage (from [1])*

| Type | | Max Drift |
|---|---|---|
| Reinforced Concrete | Shear Walls | $4.10^{-3}$ |
| | Beams and columns | $3.10^{-3}$ |
| Masonry | Unreinforced | $10^{-3}$ |

As the model remains simple, many runs can be computed and particularly many ground motions can be used as input. Therefore, the probability of (slight) damage with respect to a ground motion parameter can be studied, i.e. the first fragility curve of the building can be drawn. It is modelled as a cumulative lognormal distribution with its median and its lognormal standard deviation.

If we consider only the first mode, a simple formula can be written to have the fragility curve median with respect to the spectral displacement at the frequency of the structure ($S_d$):

$$S_{d,D1} = \frac{D_{D1}}{p_1 \max(\frac{d\Phi_1}{dx})} \quad (4)$$

with $D_{D1}$ the interstory drift limit (table 1), $p_1$ the participation factor of the first mode and $\frac{d\Phi_1}{dx}$ the space derivative of the first modal shape. The lognormal standard deviation is arbitrary set to 0.6 by comparison with the literature [1].

## Application to Grenoble and Pointe-à-Pitre

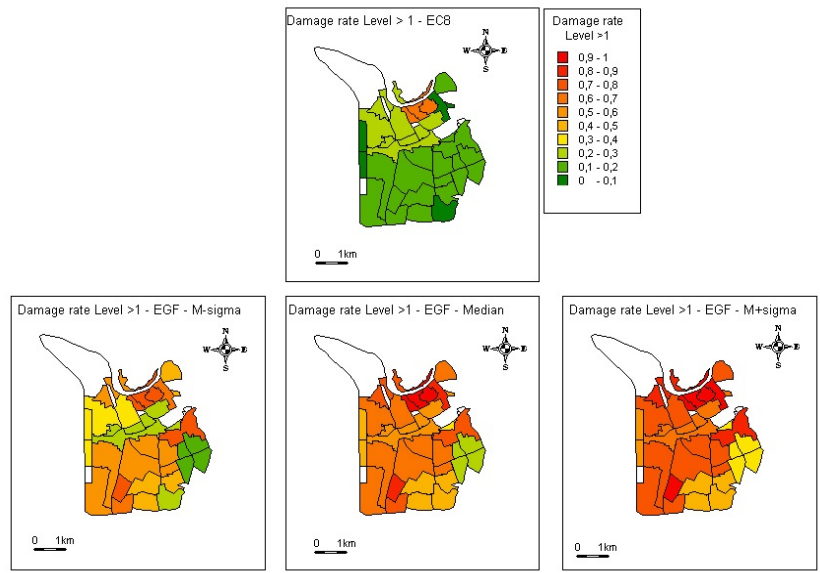

*Figure 2 – Seismic scenario (rate of buildings damaged at least slightly) for Grenoble City. Top: Design code. Bottom: Simulated earthquake including site effects with uncertainties.*

In Grenoble (France), 61 buildings of various types were tested by simultaneous recordings of ambient vibrations at least at each floor. The vulnerability of median models of each type have been assessed following the method presented above. Different seismic scenarios have been processed (Fig. 2). A scenario based on the current design code ($a_G$=1.6 m/s$^2$) showed that except in the city centre, made of medium-rise stone masonry buildings, not much damage could occur. These results are coherent with empirical vulnerability analysis [9]. However, another scenario (simulated using Empirical Green's Functions method [10]) taking site effects into account showed important damage rate in the whole city. Indeed, site effects occur at particular frequencies so that a good knowledge of the building frequencies is necessary to perform such scenarios with relevancy. That is the major advantage of the presented method.

In Pointe-à-Pitre (Guadeloupe, France), ambient vibrations of 7 buildings of the same type where recorded (3 to 16 stories). These buildings are shear wall structures built since the 1960s with the tunnel formwork technique, wide used in France at that time. Therefore, they do not have any load bearing system in the longitudinal direction that makes them very vulnerable to earthquake motion.

These buildings were subjected to the Les Saintes (Mw=6.3) and Martinique (Mw=7.4) earthquakes without damage. The accelerometric stations of the French Accelerometric Network [11] recorded these earthquakes in the city of Pointe-à-Pitre so that it has been possible to load the building models with the accelerometric signal (IPTA station). The Martinique earthquake happened to cause a greater displacement on these buildings compared to the Les Saintes earthquake (Table 2) but the calculations exhibit a low probability of damage for the 2 earthquakes (less than 1%, not displayed on the table). On the opposite, a scenario based on the current design code (EC8 spectrum, $a_G$=3.0 m/s$^2$) shows probability of damage of 40 to 60% for 5 buildings on 7 in their longitudinal direction. For another one, the probability is lower (20%) thanks to its foundation on the hard rock (soil A) whereas all the others are on soil E. The less risky building only shows a damage probability of 10% because it has only 3 stories and some longitudinal walls, therefore a higher frequency.

*Table 2 – Synthesis of the results for Pointe-à-Pitre buildings*

| Building (# stories) | f (Hz) | Les Saintes EQ | Martinique EQ | Design code | |
|---|---|---|---|---|---|
| | | $S_d$ (cm) | $S_d$ (cm) | $S_d$ (cm) | $P_{D1}$ (%) |
| Baimbridge A – L (4) | 3.36±0.01 | 0.15 | 0.33 | 2.38 | 41 |
| Baimbridge A – T (4) | 5.3±0.1 | 0.08 | 0.14 | 0.950 | 2 |
| Baimbridge B – L (4) | 3.13±0.01 | 0.17 | 0.34 | 2.76 | 54 |
| Baimbridge B – T (4) | 5.5±0.1 | 0.06 | 0.12 | 0.85 | 2 |
| Bergevin – L (3) | 4.3±0.1 | 0.13 | 0.23 | 1.42 | 53 |
| Cap. Lafitte – L (6) | 3.53±0.02 | 0.14 | 0.34 | 1.52 | 19 |
| Les Esses – L (4) | 3.5±0.1 | 0.14 | 0.34 | 2.21 | 41 |
| Les Esses – T (4) | 4.6±0.1 | 0.09 | 0.17 | 1.23 | 9 |
| Petit Pérou – L (3) | 5.1±0.1 | 0.10 | 0.16 | 1.02 | 10 |
| Petit Pérou – T (3) | 7.9±0.3 | 0.03 | 0.04 | 0.40 | 0 |
| Beauperthuis – L (16) | 1.73±0.03 | 0.46 | 0.44 | 7.73 | 56 |
| Beauperthuis – T (16) | 1.55±0.03 | 0.59 | 0.78 | 8.63 | 41 |

## Discussion and conclusions

This method is one of the first attempts to use experimental modal parameters obtained under ambient vibrations in earthquake engineering issues. The Grenoble example showed how a good knowledge of the existing buildings frequency was important in order to take into account a hazard including site effects. Hazard microzonations are useless if the resonance frequencies of buildings are not perfectly known.

The Slight damage defined in HAZUS [1] is much higher than the damage grade 1 in EMS-98 [12]. One can wonder if the experimental frequencies are still valid at this level of shaking. A method is currently developed in order to take the frequency decrease into account but still, the frequency decrease remains low, about 20% maximum [4]. This uncertainty is rather low compared to the uncertainties associated to structural analysis or empirical formulas in the design codes.

For moderate seismicity countries, this method is able to discriminate buildings that should not be reinforced and buildings that should be more precisely studied. For example, in the case of Pointe-à-Pitre, the building Petit Pérou should stay in place whereas particular care should be taken on the other buildings. Moreover, even if the Slight damage grade is not interesting for life safety issues and catastrophic earthquakes, it is valuable information for moderate earthquakes that compose the major part of the risk in North-Western Europe. In order to better estimate the financial amount of this risk, a good knowledge of the low damage grades is necessary.

The influences of the modal shape and the modal damping have not been studied, but this last parameter is also crucial (less than the frequency) for amplitude estimation. 3D ambient vibration recordings are also able to give the torsional motion of buildings so that it could also be taken into account in the modelling.

The probabilistic approach using the fragility curve concept like HAZUS [1] or other studies in the literature is of particular importance in the way that it allows to estimate the uncertainties on the risk estimation. Even if they are huge, seismic risk methods should now emphasize their estimation to show that the knowledge is progressing and to point out the parameters and methods that still need to be improved.